# Extracting Value from Grey Literature: processes and technologies for aggregating and analyzing the hidden "big data" treasure of organizations

**Gabriele Motta, Roberto Puccinelli, and Lisa Reggiani,** Network and Information System Office, CNR;
**Massimiliano Saccone,** Central Library, CNR, Italy

***Abstract***
Grey literature can be a valuable source of information about organizations' activities and, to a certain extent, about their identity. Some of the major problems that hinder their full exploitation are the heterogeneity of formats, the lack of structure, the unpredictability of their content, the size of the document bases, which can quickly become huge.
The collection and mining of grey literature can be applied to individual organizations or classes of organizations, thus enabling the analysis of the trends in particular fields. To this end, some techniques can be inherited from the best practices for the management of structured documents belonging to well identified categories, but something more is needed in our case.
Obvious steps are: identifying sources, collecting items, cleansing and de-duplicating contents, assigning unique and persistent identifiers, adding metadata and augmenting the information using other sources. These phases are common to all digital libraries but further steps are required, in our opinion, in the case of grey literature in order to build document bases of value. In particular, we think that an iterative approach would be the most suitable in this context, one including an assessment of what has been collected in order to identify possible gaps and start over with the collection phase.
We think that big data technologies, together with information retrieval and data and text mining techniques, will play a key role in this sector. This "bag of tools" will certainly facilitate the management, browsing and exploitation of large document bases that belong not only to a single organisation but also, for example, to a large number of organizations working in a particular sector. This on the one hand opens new scenarios regarding the type of information that can be extracted, but on the other hand introduces new problems regarding the homogenization of contents, formats and metadata, and additional issues related to quality control and confidentiality protection. We believe that in this context the incremental-iterative approach would help address, gradually and based on real cases, the problems mentioned above.
In this paper we describe the process that, in our opinion, should be put in place and a high level ICT architecture for its dematerialization, along with the technologies that could be leveraged for its implementation.

## 1. Introduction

Heterogeneity, complexity and structural variability characterise the 'Big data' concept and related contexts, as well as the grey literature system and its environment. It seems then natural to leverage 'Big data' technologies not only to manage grey literature contents and metadata, but also to analyse them in order to extract useful and non-trivial information, which may be hidden in them. Current 'Big data' technologies, in fact, satisfy base requirements such as scalability, flexibility (e.g. ability to cope with different formats), interoperability, openness.
On the methodological side, it seems to us appropriate to use a practical approach (i.e. non theoretical), based on some key elements:

- sharing of a fundamental language;
- collaboration between actors and communities involved;
- interoperability framework.

In the following we present our view about the analysis of grey literature contents and metadata by means of 'Big data' solutions. We start by discussing the issues related to providing a formal definition of 'Big data' and then present a short survey of software platforms and tools. We then highlight the connections between 'Big data' and grey literature and illustrate a methodological approach for combining the two concepts. We finally describe our proposal of an integrated system for the management and analysis of grey literature contents and metadata.

## 2. Big data: definition and related issues

### 2.1. Original heterogeneity and massive diffusion

The term "big data" has emerged from activities in different fields and has over time become more and more ambiguous mainly due to:

- *heterogeneity as original pattern* - the term derived from different contexts (research, industry and communication media in general) [1];
- *pervasive diffusion in many different usage contexts* - it has massively spread in very different contexts (society and public sector, business & research); examples of big data may indeed include social media data, e-business data, linked data, web usage data, government open data, research data, as well as data created in such diverse environments as cloud-based computing infrastructures, virtual collaborations and projects, e-science, e-humanities and e-social sciences.

Big data have become a cultural shift in data processing and also the preferred solution for extracting the maximum benefit from data made available by extremely heterogeneous sources. The related technologies seem to cope effectively with the information overload typical of our era [2].
Moreover, the growing usage of insights generated through Big Data Analytics in companies, universities and institutions has modified decision-making culture as well as scientific method [3].
Pervasiveness and intrinsic vague polysemy have produced a **strong semantic interference**: terms more or less synonymous for 'big data' have contaminated it at the connotative level and sometimes at the denotative level, enriching it with additional meanings or favouring its use in hybrid and different ways, so that they have been legitimated coming into common use. There are indeed many different terms used in literature that may refer to or be associated with the phenomenon of 'big data', including such terms as *digital data, research data, linked data, open data, web of data and data repositories* [2].
The multiple meanings attributed to the 'big data' concept are connected to the different production and use contexts, to the involved actors and their purposes. These definitions also show a substantial and significant diachronic dimension, changing over time in accordance with the cultural and technological evolution that characterizes application fields and user communities.
Attempts to provide operational definitions – dependent on actors, goals and circumstances and gradually refining themselves – have proliferated. They are characterized by a certain "methodical negation": approaches, methods, tools and technologies used until now would be, according to those definitions, inadequate to exploit such data whose volume is too big, whose production rate is too high and which are too heterogeneous and complex [4].

### *2.2. Big is a moving target[1]: the 3 "V"s*

The three characteristics that recur in all definitions are:

- *volume (scalability)*: the size in this case does not refer to absolute values, but is related to contexts, tools and to intrinsic constraints imposed by human-computer interaction;
- *velocity (timeliness)*: i.e. the speed at which data is generated, retrieved, and need to be processed; it is a relative concept constantly evolving, too;
- *variety*: i.e. heterogeneity - of sources, types, internal structures, formats, etc. - and complexity, linked to different contexts, instruments, procedures and purposes of production, collection, management and reuse of data.

Most of the frequently mentioned definitions include and/or refer to these three characteristics, starting from a MetaGroup (now Gartner) report (2001). The Gartner report made no mention of the term "big data" and exclusively used the general term "data" with reference to data management and data management challenges, emphasizing the relationship with e-commerce. It repeatedly alluded to huge and continuously growing amounts of data and highlighted the increasing size of data, the increasing rate at which they are produced and the increasing range of adopted formats. Gartner proposed a threefold definition encompassing the "three Vs":

[1] Nature Special Issue, 4th September 2008 - http://www.nature.com/news/specials/bigdata/index.html (focusing on big data and their implications for science) [5].

*Volume, Velocity, Variety*, predating the current trend [6]. The three features, crystallizing over time, are always associated – in whole or in part – with the 'big data' notion.
Given the high variability and volatility of the data produced in many fields of interest, a fourth feature has progressively gained a significant relevance: veracity, a characteristic required to provide a solid ground for information extraction and effective data reuse. It is tightly tied to quality of data, data sources and processes – concepts that can be summed up by the terms 'accuracy', 'reliability', 'traceability' and 'trustability'. A recent survey on big data (2015) provided the following definition:
"Veracity (Data in doubt). Veracity is what is conform with truth or fact, or in short, Accuracy, Certainty, Precision. Uncertainty can be caused by inconsistencies, model approximations, ambiguities, deception, fraud, duplication, incompleteness, spam and latency. Due to veracity, results derived from Big data cannot be proven; but they can be assigned a probability [7]."
Therefore, within the four characteristics above described preferential relationships can be highlighted between:

- volume (scalability) & velocity (timeliness): *quantitative*, i.e. measurable aspects, apparently objective and thus more easily definable and manageable;
- variety & veracity/validity: *qualitative*, i.e. not measurable aspects, difficult to control, to define and to use [4].

The stress on the first two properties derives, at least partially, from the original inception and use of big data within business and commerce, hard sciences and ICTs sectors. This stress has not only persisted but has also increased over time: technologies developed for data analysis as well as control of data useful to identify choices, habits and behaviours of consumers/users have predictably focused on big data quantitative measurable dimensions, that continue to play a critical role in the current business models.
To sum up, the 'big data' concept consists of: data, relation contexts, technologies and tools that ensure their use.

## 3. Big data and Technologies

This section presents a brief survey of the most widely used Big Data technologies. For further details, see [16] and [17].

### 3.1. Big Data storage and processing technologies

As regards "Big Data" storage and processing, the first word that comes to mind to insiders is "Hadoop". This is an Apache project mainly focused on developing four modules:

- *Hadoop Common*: the common utilities that support the other Hadoop modules.
- *Hadoop Distributed File System (HDFS™)*: a distributed file system that provides high-throughput access to application data.
- *Hadoop YARN*: a framework for job scheduling and cluster resource management.
- *Hadoop MapReduce*: a YARN-based system for parallel processing of large data sets.

The framework is written in Java and its goal is to provide the ability to store and process large amounts of data using commodity hardware. One of its most notable features is to provide high availability by detecting and handling failures at the application layer, thus freeing from the need for complex hardware-based solutions.
For the purposes of this paper, the two modules of interest are HDFS and MapReduce, which are described in detail below.

HDFS. It is a distributed file system designed in reference to Google File System (GFS) in order to run on large clusters of commodity hardware. As described in [16] "HDFS is the basis for main data storage of Hadoop applications, which distributes files in data blocks of 64MB and stores such data blocks in different nodes of a cluster, so as to enable parallel computing for MapReduce. An HDFS cluster includes a single NameNode for managing the metadata of the file system and DataNodes for storing actual data. A file is divided into one or multiple blocks and such blocks are stored in DataNodes. Copies of blocks are distributed to different DataNodes to prevent data loss." The HDFS interface has been designed based on a similar approach to that of the Unix file system.
*MapReduce*. This framework implements the homonymous programming model, which is currently one of the most widely used for processing large datasets. It includes the ResourceManger module, running on the master node of the cluster and orchestrating the computational resources, the NodeManager module, running on the slave nodes (one instance per node) that perform the actual work, and the MRappMaster module (one per application). A MapReduce job handles input and output data sets in the form of <key, value> pairs, typically stored in a distributed file-system. The MapReduce framework and the HDFS generally run on the same set of nodes, thus enabling to dispatch jobs to the nodes where data reside. This approach reduces the bandwidth usage.
In the MapReduce paradigm, the Map function provides intermediate outputs as sets of <key, value> pairs and the Reduce function merges the intermediate values with the same key.
As stated in the official MapReduce tutorial "… Minimally, Hadoop applications specify the input/output locations and supply map and reduce functions via implementations of appropriate interfaces and/or abstract-classes. These, and other job parameters, comprise the job configuration. The Hadoop job client then submits the job (jar/executable etc.) and configuration to the ResourceManager which then assumes the responsibility of distributing the software/configuration to the slaves, scheduling tasks and monitoring them, providing status and diagnostic information to the job-client."[2].

------

It is worth mentioning that MapReduce is not the only Big Data processing paradigm. Other programming models are [16]:

- *Dryad*, whose operational structure is represented as a directed acyclic graph, in which vertexes represent programs and edges represent data channels;
- *All pairs*, that focuses on comparing element pairs in two datasets by a given function;
- *Pregel*, where each computational task is expressed by a directed graph constituted by vertexes and directed edges, in which every vertex is related to a modifiable and user-defined value and every edge is related to its source vertex and is constituted by a modifiable and user-defined value and an identifier of a target vertex.

### 3.2. NoSQL Data bases

The term "NoSQL" generically refers to database technologies not based on the relational model. It is by some sources presented as an acronym standing for "Not only SQL", thus implying that, in this context, the use of the SQL is neither excluded nor mandatory.
This set of technologies has been developed to overcome the limits of the relational model, mainly in terms of horizontal scalability and/or performance. There are several types of NoSQL databases, each one focused on or more suited for a particular set of problems. A particular NoSQL database may be faster than an RDBMS for some operations and slower for other operations.
We do not present in the following a complete survey of NoSQL databases but just provide a list of the main types, each one with a short list of examples.

### 3.3. Key value databases

In this type of data base, data are stored in the form of <key, value> pairs. The underlying data model is the associative array, i.e. a set of <key, value> pairs where keys are unique. Pairs can be ordered lexicographically to improve performance of interval queries and values can be retrieved based on the corresponding keys. Examples of KV stores are :
*Amazon Dynamo*, a highly available and expandable distributed key-value data storage with a simple read-write interface. Flexibility and high-availability are achieved through data partitioning and replication. Consistency across replicas is guaranteed as an asymptotic condition (eventual consistency);
*Voldemort*, developed for LinkedIn, is a KV store where key words and values can be composite objects constituted by tables and images. Its interface includes three simple operations: reading, writing, and deletion. Voldemort does not ensure data consistency but supports optimistic locking for consistent multi-record updating. Data can be managed in RAM or via storage engine (Berkeley DB and Random Access Files).
Other examples of KV stores are Redis, Tokyo Canbinet and Tokyo Tyrant, Memcached and MemcacheDB, Riak and Scalaris.

### 3.4. Column-oriented databases

As suggested by the name, databases of this type store and process data according to columns other than rows. Columns and rows are segmented in multiple nodes to realize expandability

[2] https://hadoop.apache.org/docs/r1.2.1/mapred_tutorial.html

[16]. In a relational database system, tables are composed by rows, each one usually identified by a primary key. Example:

Employee table:

| RowID | First Name | Middle Name | Last Name | Department |
|---|---|---|---|---|
| 0001 | Michael | John | Stanford | Sales |
| 0002 | Mary | Ann | Naughton | R&D |
| 0003 | Michael | Bradley | Rutherford | Operations |
| 0004 | David | Robert | Jones | R&D |

In a columnar database the above table could be stored as:

Michael: 0001, 3; Mary: 0002; David: 0004;
John: 0001; Ann: 0002; Bradley: 0003; Robert: 0004;
Stanford: 0001; Naughton: 0002; Rutherford: 0003; Jones: 0004;
Sales: 0001; R&D: 0002, 0004; Operations:0003;

Using this model, queries like “find all employees with First name ‘Michael’” would be performed with a single access. In addition, aggregated queries, like the ones usually performed in data warehouses, can benefit from this type of layout.

### 3.5. Document Databases

Databases of this type store records as documents encoded according to several standards (JSON, BSON, XML, YAML, etc.). Each document is identified by a unique key. They generally allow for data distribution and redundancy over clusters of servers in order to achieve horizontal scalability, high availability and fault tolerance.
Examples of document database are MongoDB, SimpleDB and CouchDB.
*MongoDB* is an open source platform where documents and queries are both represented as BSON and query syntax is similar to JSON. Automatic sharding distributes data across a cluster of machines. Eventual consistency is guaranteed and Multi-Version-Concurrency-Control can be implemented on top of the MongoDB features.
*SimpleDB* is an Amazon Service accessible via web service API. It is a distributed database but does not support automatic partition. It allows users to use SQL to run query. Eventual consistency is guaranteed but not MVCC.
*Apache CouchDB* is an open source platform where data are stored as JSON. It features a RESTful HTTP API for querying and updating documents. Each document is uniquely identified. “If a document needs to be modified, the client can download the entire document, modify it, and then send it back to the database. After a document is rewritten once, the identifier will be modified and updated”. [vd fonte] CouchDB features eventual consistency and implements synchronous MVCC. ACID The CouchDB file layout and commitment system also features all Atomic Consistent Isolated Durable (ACID) properties. It also features a view model that allow aggregation and reporting on the documents in the database. Views are defined via Javascript as a Map function in a MapReduce system.

### 3.6. Triple stores

A triple store keeps information in the form of (subject, predicate, object) triples and is generally suited for semantic applications. Triples can be represented in RDF (Resource Description Framework) format and data can be retrieved via a query language (e.g. SPARQL). Examples of triple stores are AllegroGraph, Apache Jena, Oracle NoSQl Database, SparkleDB, Virtuoso Universal Server.

### 3.7. Other types of NoSQL databases

Other types of NoSQL databases are Graph databases, Object databases, Tabular databases, Tuple stores.

### 3.8. Data warehousing

When it comes to heterogeneous data sources, one possible approach to the collection and analysis is to apply adequate procedures that extract data from the original repositories, process them according to the needs, and finally, load them into appropriate structures, optimized for aggregated queries (ETL procedures). Data warehousing tools in the Big Data field are:

- *Hive*, a software that allows to project a unified structure onto distributed heterogeneous data and to query them using an SQL like language;
- *Pig*, a platform that allows to compile and run MapReduce programs written in a textual

language called Pig Latin;
- *Sqoop*, a tool designed for efficiently transferring data between structured, semi-structured and unstructured data sources.

### 3.9. Big Data Analytics tools

In 2012 KDNugget conducted a survey among 798 professional asking the following question:
"What Analytics, Data mining, Big Data software you used in the past 12 months for a real project?"
The top five answers were [16]:
- *R*, an open source execution environment for programs written in the homonymous language;
- *Excel*, the only commercial software among the top five;
- *Rapid-I Rapidminer*, a tool that allow to define and execute complex analysis jobs described as xml files and visualized through an intuitive GUI;
- *Knime*, written in Java and based on the Eclipse framework, features a visual GUI and can be easily extended via plugins;
- *Weka*, written in Java and integrated with the Pentaho BI platform, provides such functions as data processing, feature selection, classification, regression, clustering, association rule, and visualization.

For the purposes of this paper, also tools for text mining that can perform document clustering and classification are of interest. Examples are:
- *Rapidminer* with its Text Processing Extension;
- *Knime* Text Processing extension;
- *Carrot2*.

## 4. Big data and Grey Literature

With regards to the grey literature big data include [8]:
- research data (in their processing stages, starting from the raw data set acquisition), textual and not-textual documents and virtual data representations;
- the contexts and their relationships, typical of the complexity of contemporary science, of its various actors and communities;
- infrastructures, instruments, tools and ICT methods.

Variety – as heterogeneity and complexity – and, more generally, structural variability that characterize big data (i.e. data, context & relationships, instruments & tools) emphasize the intrinsic critical features of grey literature.

### 4.1. Benefits and hidden value: the synergy between big data and grey literature

The *quality* dimension in its articulation – above highlighted for big data – has become decisive in grey literature and in its connection with big data, because on the one hand it enables to increase and fully exploit grey literature potential and on the other it allows to effectively cope with actual shortcomings and risks. In fact, it drives the various stakeholders and communities to improve all the steps of process, (acquisition, processing, dissemination) and to focus in particular on data and metadata quality control [9].
Taking into account the big data characteristics and adequately valuing their contexts and relationships, it is possible to:
- improve communication between actors and communities;
- discover more easily implicit research trends and/or latent themes, both at the individual and community level, especially inter-, trans- and multi-disciplinary ones[10];
- foster multi-, trans- and inter-disciplinary research;
- create and strengthen effective antidotes – approaches, methods and practices – to hyper-specialization, that always means 'non-communication' and mutual incomprehension [11].

The discovery of hidden research trends and latent topics [11] can be useful to various actors, in particular:
- to researchers, to identify mainstream research trends, before they emerge to the surface, and to detect upcoming/implicit research trends or interesting connections between research groups/fields;

- to policy makers, because it supports them in long term research investment planning[3];
- to industries, because it anticipates innovation, fostering it; thereby it accelerates and facilitates convergence of research and innovation, a key trend in the current national and European planning.

## 5. Methodological approach

From a methodological perspective, in order to take the most advantage from grey literature and big data interaction, it is necessary to focus on some interconnected aspects of quality:

- transparency in methodology, experimental observation, and data collection;
- reliability and reusability of scientific data and research output;
- accessibility and transparency of scientific communication processes;
- adoption of web-based tools to facilitate scientific collaboration and to enhance process accessibility and transparency.

### 5.1. Context and reference framework

In order to achieve the above listed goals it is necessary to define:

- the conceptual framework and methodology;
- the fundamental processes involved, with particular attention to those of scholarly communication;
- a common language among stakeholders and research communities that can go beyond peculiarities of individual entities (institutions/companies/persons) and groups (disciplinary and/or national communities, teams, research networks and communities…).

The *Reference Framework* we propose for Big Data and scholarly communication (which includes Grey Literature) is complex and heterogeneous, and presents different aspects:

- multidimensionality: it regards different sectors and settings of intervention, which are developed at different levels;
- interdisciplinarity and multidisciplinarity;
- diachronicity: it characterises the influential factors and their mutual interdependencies.

### 5.2. A methodological road map

The above listed aspects should, in our opinion, be addressed by:

- identifying the different stakeholders (researchers, policy makers, ICT and information specialists, citizen, industries, libraries, etc.), and their related purposes and strategies;
- identifying and knowing the different information and cultural contexts, practices, approaches, channels and specific jargons (technical jargon);
- identifying and using consistently the instruments/tools that allow to plan and manage the actions in the R&D system at multiple levels and in different intervention areas (International, European and National policies and programmes, guidelines; laws & rules; institutional regulations; agreements, contracts, etc.);
- verifying implementation of the interventions and their effectiveness at different levels, through monitoring and evaluation systems;
- as a consequence, guiding the choices on strategies, instruments, actions at different levels and environments on the basis of the analysis previously carried out (by confronting and relating results).

### 5.3. Interoperability

In order to act on processes, it is necessary to define a common interoperability framework, i.e. to define a common multi-level and multi-dimensional framework in which main stakeholders and different research communities can share objectives, resources, infrastructures, instruments/tools, services and results.

The strategic intervention areas are the following: political, institutional and economic setting; legislative, legal and regulatory framework; organizational setting; technical and technological environment.

However, firstly, culture, meaning and language sharing is the basis of all the interoperability strategies. In fact, this approach preliminarily requires the use of a common language among the different contexts and actors, that is, a common cultural code.

---

[3] Big data usage requires that stakeholders are aware of all the legal and ethics-related issues (privacy, security, intellectual property and copyright, etc.) [12].

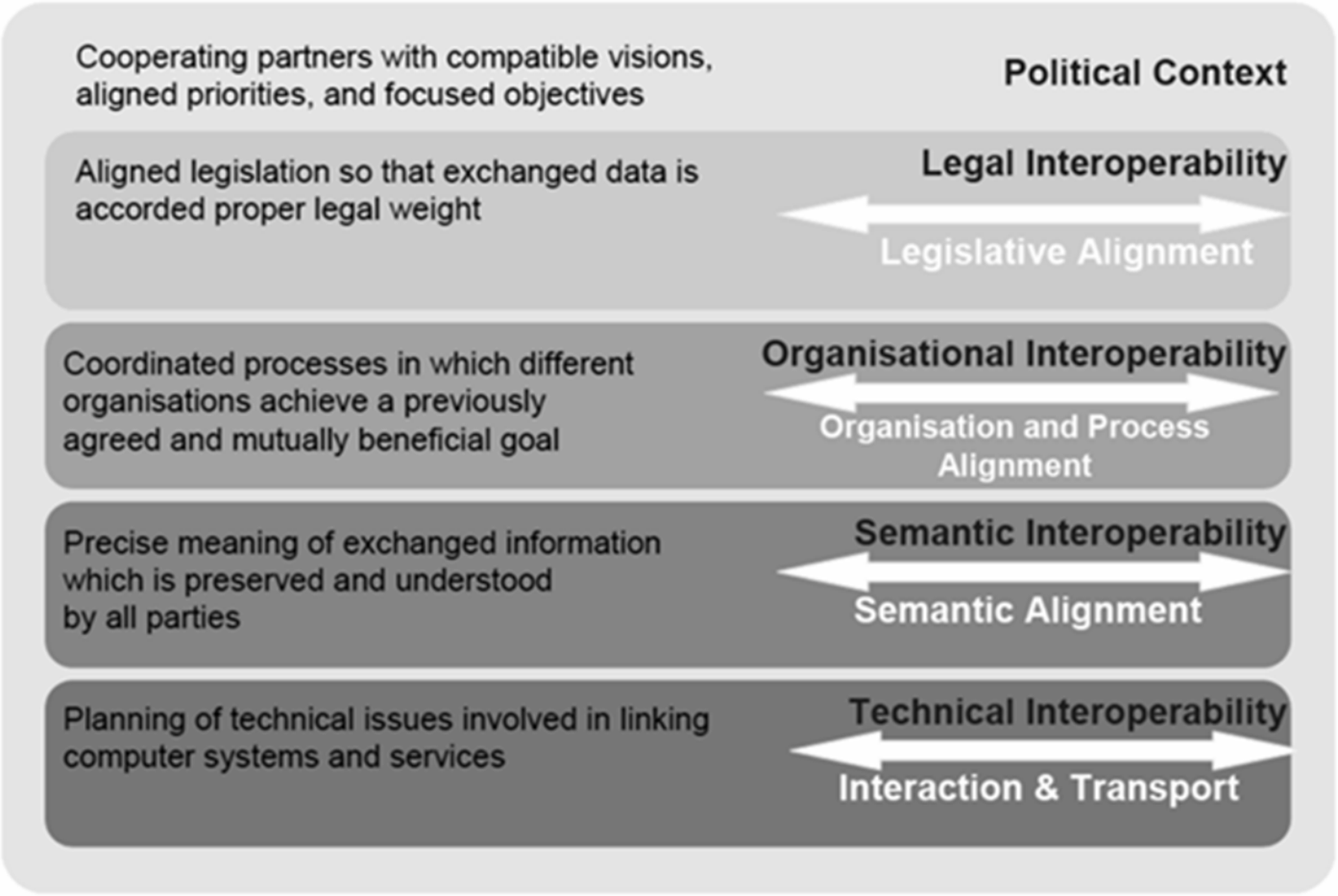

Figure 1 - European Interoperability Framework (EIF) [13]

### 5.4. A first proposal

The different issues, quickly examined in the previous section, are addressed by a first methodological proposal, which includes three sequential steps:

1. to address key challenges, regarding the entire big data life cycle and their production, collection, management and re-use systems, through a multi-dimensional analysis, that methodically involves different communities and actors;
2. to build a stable collaboration between actors and communities, characterized by very different cultures and interests, often competing with each other (i.e. to understand complexity and diversity in order to maximize the heterogeneity potential);
3. to exploit Linked Open Data as context of relationships, in order to promote the exchange of good practices and to improve the transparency of processes.

## *6. Proposed technical solutions*

In the following sections, we illustrate our proposal of an integrated system for the collection, processing, analysis and dissemination of grey literature contents coming from different communities. Such system, in our opinion, allows extracting useful information from the collected data and metadata, thus allowing the full exploitation of their potential. We start from the analysis types we deem interesting for our purposes and then present a high level logical view of the architecture. Finally, we illustrate the workflow that should be implemented in order to acquire, cleanse, enrich and process the items of interest.

### 6.1. Analysis types

There are many types of analysis that can be performed on a document base of the kind addressed in this paper. In order to focus our activities on the needs of the stakeholders listed in the previous paragraphs (political decision makers, industry, researchers), we decided to initially restrict our attention to a limited number of analysis, whose potential benefits are more obvious. In first place, we think that the heterogeneity of document types, formats and sources poses serious problems of classification and in general of metadata completeness. To this end, we think that applying well-selected classification techniques might be of help. Machine learning algorithms of this type take as input:

- A document base;
- A list of categories;
- A training set (i.e. a set of documents for which the classification has already been performed and is assumed to be correct);

and assign each document to one category. This process may be used to fill some metadata fields whose values are missing.

Another interesting use of machine learning algorithms could be the application of clustering techniques for the detection of non-trivial connections between research fields or organizations/research groups. Let us assume to have partitioned a document base into clusters based on the textual content. We could then evaluate the relative frequencies of the values of particular metadata fields within each cluster in order to identify patterns and/or correlations.

As a last example, we would like to propose the application of association rules to detect interesting patterns. When applied to the grey literature, these may be useful for the early detection of underlying trends that may not (yet) be evident from journal articles or conference proceedings papers. As a result of such analysis we may, for instance, discover that, if a research group is specialized in field A and is based in the geographic area B, then they have a strong probability to participate to initiatives in the field C. For further details about the above-described algorithms, see also [14] and [15].
As a final remark, we would like to point out that analyses may be performed both on metadata and contents and that it is also important in a system as the one we are proposing to have classical Business Intelligence tools in order to perform normal aggregate and detail reporting.

### 6.2. Proposed architecture

In this section, we present a high level logical view of a system for the acquisition and analysis of grey literature contents. This system is designed to serve different user communities and to acquire contents in different formats from providers who have signed agreements that bind them to respect some basic requirements in terms of content quality. It is a multi-layer architecture featuring the “Data layer” at its base. This is the level where contents, metadata and analysis results are kept, each one in the appropriate type of storage. We think that contents should be stored in a distributed file system whereas metadata are better managed using a NoSQL database, (e.g. an eventually consistent Key-Value Store like Cassandra). Analysis results may be stored either in a NoSQL database or in a relational one. A triple store is included for semantic data.
On top of the “Data layer”, there is the “Access layer”, which provides access to metadata and contents in both input (Ingestion) and output (Retrieval). As regards ingestion, a high throughput, multi-threaded, queue-based collection system is required, as the one developed for the “Science and Technology Digital Library” project [18], successfully completed by AgID and CNR. Such a component feeds a cleansing and enrichment workflow that prepares metadata and contents for the analysis. As regards retrieval, search functions are implemented at this level using components like Apache Solr. Search and browsing functions are exposed to the upper layer as RESTful API.
The “Service layer” includes components for text and data mining as the ones presented in the previous sections. In addition, classic BI tools are included for normal reporting. At the top of the stack is the Presentation layer, which includes portals for the access of contents, analysis results and reports.
As we said before, different user communities may access this system, thus requiring access control mechanisms and user profiling functions.

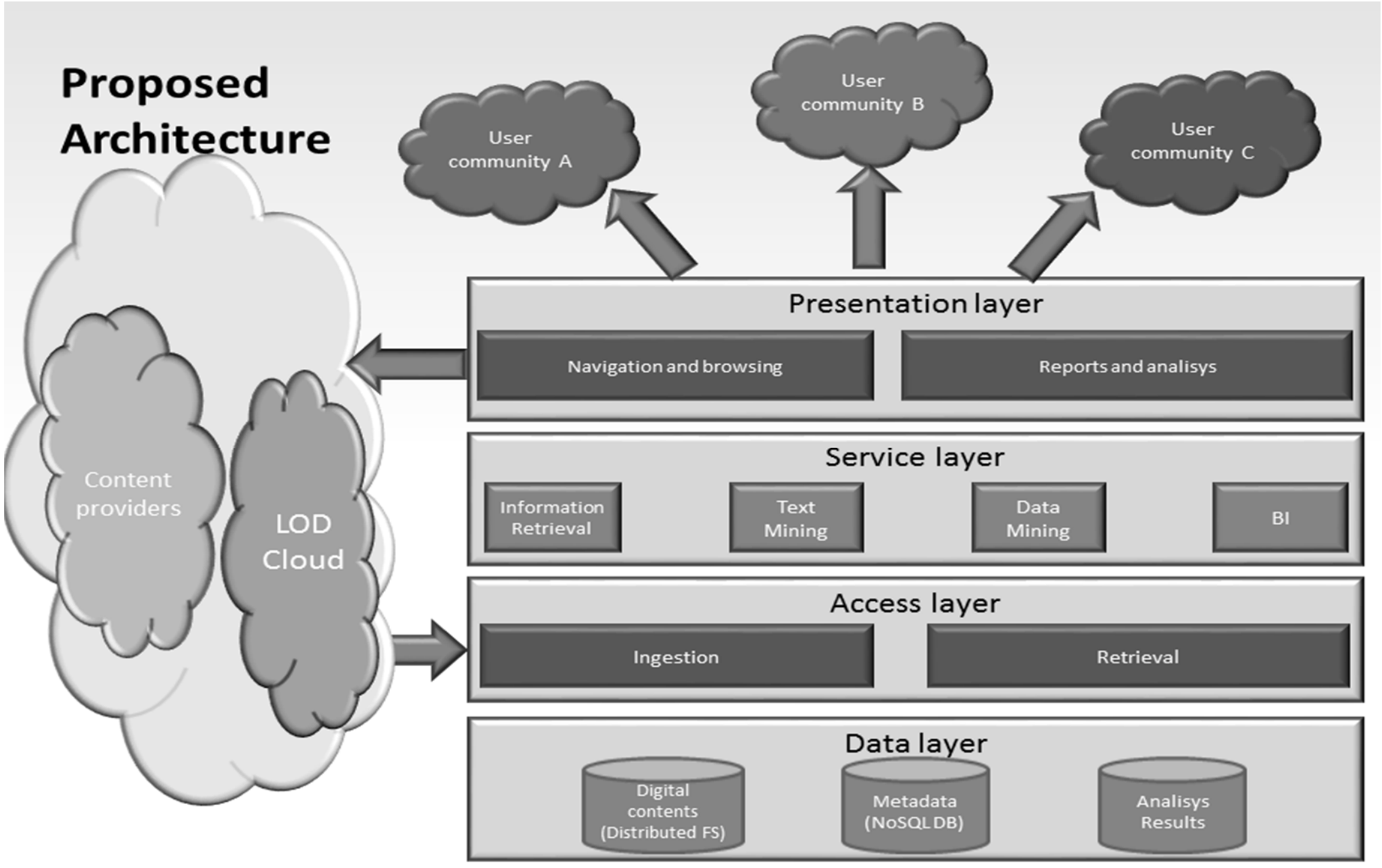

Figure 2 - Proposed architecture (logical view)

### 6.3. Proposed workflows

As regards the ingestion and processing workflow, contents are submitted to the system as Submission Information Packages (SIP). For each SIP, contents and metadata are extracted. Metadata are cleansed and enriched using inference algorithms, external sources and content-bundled metadata (i.e. metadata enclosed in the content files as in PDF, Word, JPEG, etc.). Finally, metadata are fed to the data mining algorithms.
On the other hand, contents pass a virus check and, in case of textual documents, a text extraction step. A deduplication algorithm helps detecting multiple copies of the same item. Contents too are fed to the data and text mining algorithms. Search indexes are periodically refreshed to ensure content retrieval.

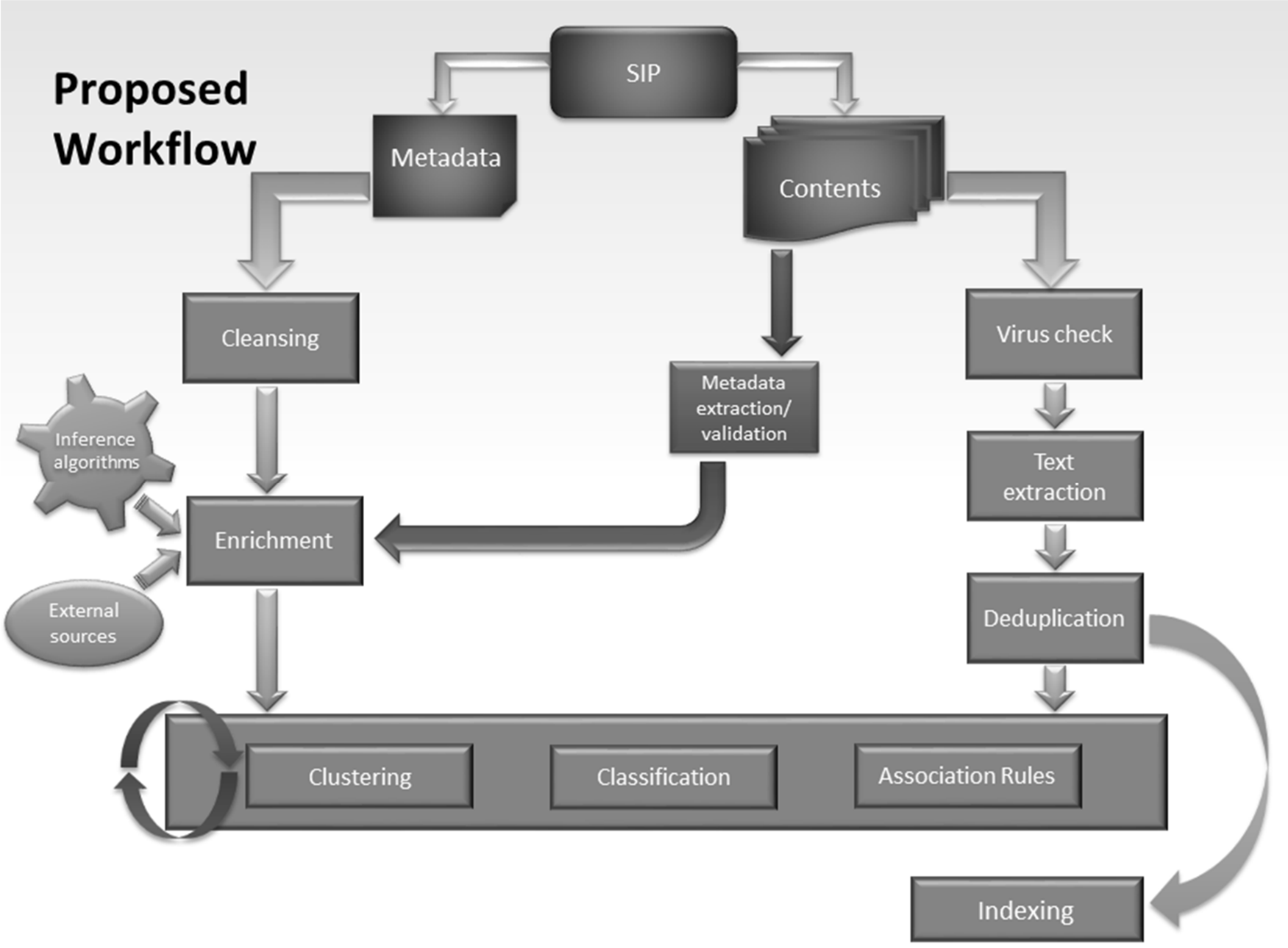

Figure 3 - Proposed core workflow

### *7. Conclusions and future work*

In this paper we have presented a methodological framework for the management, processing and analysis of grey literature contents and metadata, along with a high-level ICT architecture based on 'Big data' technologies. The methodological framework can, in our opinion, be useful to tackle the main issues regarding the addressed problems, mostly related to organisational and communication aspects, and in particular to cope with interoperability concerns.
It can also contribute to effectively promote the network of relationships among the main actors and communities.
This approach, in our view, favours a) the integration of metadata, contents and analysis results in the LOD cloud and b) the exploitation of feedback from user communities for the continuous improvement of analysis result quality.
Moreover, despite the inherent differences of cultural approaches, it makes possible the gradual convergence of the involved actors about some key definitions and related evaluation criteria.
On the technical side, one important issue that will be addressed in our future activities regards the criteria for the selection of contents and metadata to be analysed. It is indeed a non-trivial task in such a heterogeneous context to ensure at the same time richness and meaningfulness of the information content and absence of noise.